# $BaT_2As_2$ Single Crystals (T = Fe, Co, Ni) and Superconductivity upon Co-doping

Athena S. Sefat*, David J. Singh*, Rongying Jin*, Michael A. McGuire*, Brian C. Sales*, Filip Ronning+, David Mandrus*

*Materials Science & Technology Division, Oak Ridge National Laboratory, Oak Ridge, TN 37831, USA**

*Los Alamos National Laboratory, Los Alamos, NM 87545, USA +*

## Abstract

The crystal structure and physical properties of $BaFe_2As_2$, $BaCo_2As_2$, and $BaNi_2As_2$ single crystals are surveyed. $BaFe_2As_2$ gives a magnetic and structural transition at $T_N$ = 132(1) K, $BaCo_2As_2$ is a paramagnetic metal, while $BaNi_2As_2$ has a structural phase transition at $T_0$ = 131 K, followed by superconductivity below $T_c$ = 0.69 K. The bulk superconductivity in Co-doped $BaFe_2As_2$ below $T_c$ = 22 K is demonstrated by resistivity, magnetic susceptibility, and specific heat data. In contrast to the cuprates, the Fe-based system appears to tolerate considerable disorder in the transition metal layers. First principles calculations for $BaFe_{1.84}Co_{0.16}As_2$ indicate the inter-band scattering due to Co is weak.

## 1. Introduction

The discovery of superconductivity in a variety of compounds with square lattice sheets of $Fe^{2+}$ in the structure has created great attention primarily due to their non-$Cu^{2+}$ based origin. The structure types adopted by iron-based superconductors include those of ZrCuSiAs-type *R*FeAsO [1, 2], and $ThCr_2Si_2$-type $BaFe_2As_2$ [3]. In the oxypnictide *R*FeAsO system, electron- and hole-doping has produced critical temperatures ($T_c$) as high as high as ≈ 55 K in $SmFeAsO_{0.9}F_{0.1}$ [4], $Gd_{1.8}Th_{0.2}FeAsO$ [5], and $SmFeAsO_{0.85}$ [6]. In the oxygen-free compounds of $AFe_2As_2$ (*A* = Ca, Sr, Ba) system, hole-doping has reached $T_c$ values of 38 K by K-doping [7]. The superconductivity induced by electron-doping with cobalt in $BaFe_2As_2$ was reported by us earlier [8]. This demonstrates that in-plane disorder is highly tolerated in Fe-sheets, without creating localized moments. This scenario is consistent with our report on LaFeAsO system [2] and in contrast with the high $T_c$ cuprates, as $Zn^{2+}$ doping induces localized moments and destroys superconductivity. Due to the existing interest in the properties of $ThCr_2Si_2$-type $BaT_2As_2$ materials (T = transition metal), the thermodynamic and transport properties of single crystalline T = Fe, Co, and Ni samples are surveyed. Also, bulk superconductivity by Co-doping in $BaFe_2As_2$ single crystals is reviewed.

## 2. Results & Discussions

In the preparation of crystals, high purity elements (> 99.9 %) were used and the source of the elements was Alfa Aesar. The single crystals of $BaT_2As_2$ were grown each out of its respective TAs binary where T = Fe, Co, or Ni [8, 9, 10]. Such self flux is preferred over other metal solvents such as Sn or Pb, as flux impurities can become incorporated in the crystals. For $BaFe_2As_2$ or $BaCo_2As_2$, a 1:5 ratio of Ba:FeAs or Ba:CoAs was heated for 15 hours at 1180°C under partial argon atmosphere. In both cases the ampoules were cooled at the rate of 2-4 °C/hour, followed by decanting of flux by use of a centrifuge at 1090 °C. For Co-doped $BaFe_2As_2$, a ratio of Ba:FeAs:CoAs = 1:4.45:0.55 was heated to 1180 °C, and held for 10 hours. The reaction was cooled rate of 3 - 4 °C/hour, followed by decanting of FeAs flux at 1090 °C. For the growth of $BaNi_2As_2$ single crystals, a ratio of Ba:NiAs = 1:4 was heated in an alumina crucible for 10 hours at 1180°C under a partial atmosphere of argon. This reaction was cooled at the rate of 3 °C/hour, followed by decanting of flux at 1025 °C. The typical crystal sizes from all batches were ~ 6 x 5 x 0.2 $mm^3$. The crystals were brittle, well-formed plates with the [001] direction perpendicular to the plane of the crystals. The $BaCo_2As_2$ and $BaNi_2As_2$ crystals were found to be highly air-sensitive. Electron probe microanalysis of a cleaved surface of the single crystal was performed on a JEOL JSM-840 scanning electron microscope using an accelerating voltage of 15 kV and a current of 20 nA with an EDAX brand energy-dispersive X-ray spectroscopy (EDS) device. EDS analyses on all parent crystals indicated Ba:T:As ratio of 1:2:2. EDS analyses on several crystals of the Co-doped $BaFe_2As_2$ indicated that 8.0(5) % of the Fe is replaced by Co in $BaFe_2As_2$. This composition is presented as $BaFe_{1.84}Co_{0.16}As_2$.

$BaT_2As_2$ (T = Fe, Co, Ni) and Co-doped $BaFe_2As_2$ crystallize with the $ThCr_2Si_2$ structure-type [11, 12] at room temperature, in tetragonal space group symmetry *I*4/*mmm* (No. 139; Z = 2). The crystal structure is made up of $Ba_{0.5}^{2+}(TAs)^-$ for the parents, with partial and random substitution of Co on Fe sites for $BaFe_{1.84}Co_{0.16}As_2$. At room temperature, the transition metal atoms lie on a perfect square net in the *ab*-plane of the tetragonal structure. The phase purity of the crystals was determined using a Scintag XDS 2000 2θ-2θ diffractometer (Cu $K_\alpha$ radiation). The cell parameters were refined using least squares fitting of the peak positions in the range 2θ from 10 - 90° using the Jade 6.1 MDI package. These are shown in Fig. 1. The relative decrease in *c*- lattice parameter is ~ 12% on going from $BaFe_2As_2$ to $BaNi_2As_2$. The increase in *a*- lattice parameter is smaller (Fig. 1). The refined lattice constants of $BaFe_{1.84}Co_{0.16}As_2$ are $a$ = 3.9639(4) Å and $c$ = 12.980(1) Å. Cobalt doping results in a small decrease (0.3 %) in the length of the $BaFe_2As_2$ *c*-axis, while the value of *a*-axis is unchanged within experimental uncertainty. We have reported similar behavior in Co-doped LaFeAsO [2].

Upon cooling, $BaFe_2As_2$ and $BaNi_2As_2$ undergo symmetry-lowering crystallographic phase transitions. For $BaFe_2As_2$, a transition at 132 K is associated with a tetragonal to orthorhombic (*Fmmm*) symmetry. Below the phase transition, the four equal Fe-Fe bonds at 280.2 pm are split into two pairs with 280.8 pm and 287.7 pm lengths [3]. For $BaNi_2As_2$, a tetragonal to orthorhombic first-order phase transition was suggested [13] below $T_0$ = 131 K only by analogy with $A$Fe$_2$As$_2$ ($A$ = Ba, Sr, Ca) compounds. However, we have recently found that the symmetry of the low-temperature structure is triclinic ($P\bar{1}$) [10]. The reduction in symmetry below $T_0$ results in a distorted Ni network with short Ni-Ni contacts (2.8 Å) and longer Ni-Ni distances (3.1 Å). For $BaCo_2As_2$, the low temperature diffraction was not studied as no structural transition is expected based on the physical properties, summarized below.

DC magnetization was measured as a function of temperature using a Quantum Design Magnetic Property Measurement System. Fig. 2a shows the measured magnetic susceptibility ($\chi$) in zero-field-cooled (*zfc*) form for $BaFe_2As_2$ in 1 Tesla along *c*- and *ab*-crystallographic directions. At room temperature, $\chi_c \approx \chi_{ab} \approx 7 \times 10^{-4}$ cm$^3$ mol$^{-1}$. The susceptibility decreases linearly with decreasing temperature, and drops abruptly below ~ 135 K, with $\chi_c > \chi_{ab}$ below. The polycrystalline average of the susceptibility data are presented as the Fisher's d($\chi$T)/dT [14] versus temperature (Fig. 2a, inset) to infer $T_N$ = 132 K. Shown in Fig. 2b is the temperature dependence of $4\pi\chi_{eff}$ measured along *ab*-plane at an applied field of 20 Oe for $BaFe_{1.84}Co_{0.16}As_2$, where $\chi_{eff}$ is the effective magnetic susceptibility after demagnetization effect correction, as described in ref. [8]. The *zfc* $4\pi\chi_{eff}$ data decreases rapidly for temperatures below ~ 22 K and quickly saturates to 1. This indicates that the system is indeed fully shielded and the substitution of Co on Fe-site is more homogenous than the use of other dopants such as fluorine in $LaFeAsO_{1-x}F_x$ [2] or potassium in $Ba_{1-x}K_xFe_2As_2$ [7] for which the transitions are broadened. In addition, cobalt is much easier to handle than alkali metals or fluorine. The *fc* $4\pi\chi_{eff}$ saturates at much smaller value (< 0.01), reflecting strong pinning due to the Co dopant. As seen in the inset of Fig. 2b, $\chi$ for $BaFe_{1.84}Co_{0.16}As_2$ varies smoothly between ~ 22 K and 300 K, indicative of the disappearance of magnetic ordering. For $BaNi_2As_2$, the magnetic susceptibility is anisotropic (Fig. 2c). $\chi$ is roughly temperature independent and presents only a small drop at $T_0 \approx 131$ K and a rise below ~ 12 K. For $BaCo_2As_2$, the magnetic susceptibility decreases with increasing temperature and is anisotropic with $\chi_{ab}/\chi_c \approx 0.7$ (Fig. 2c, inset). At 1.8 K, $\chi_c = 5.4 \times 10^{-3}$ mol$^{-1}$Oe$^{-1}$ and $\chi_{ab} = 3.8 \times 10^{-3}$ mol$^{-1}$Oe$^{-1}$. The field dependent magnetization for T = Ni and Co are linear at 1.8 K (data not shown).

Temperature dependent electrical resistivity was performed on a Quantum Design Physical Property Measurement System (PPMS). The resistivity was measured in the *ab*-plane ($\rho_{ab}$), as described in ref. [8]. While both $BaFe_2As_2$ and $BaFe_{1.84}Co_{0.16}As_2$ show metallic behavior, the resistivity for $BaFe_{1.84}Co_{0.16}As_2$ is smaller than for the parent compound (Fig. 3a). At room temperature, $\rho_{ab}$ = 0.50 m$\Omega$ cm for $BaFe_2As_2$ and $\rho_{ab}$ = 0.32 m$\Omega$ cm for $BaFe_{1.84}Co_{0.16}As_2$. For $BaFe_2As_2$ below ~135 K, there is a sharp step-like drop. This feature is best manifested in the derivative of resistivity, d$\rho$/dT, giving a peak at 132(1) K. For $BaFe_2As_2$ the residual resistivity ratio RRR (= $\rho_{300K}/\rho_{2K}$) is 4. The magnitude of the resistivity at 2 K and 8 Tesla is higher than the zero-field value for $BaFe_2As_2$ (top inset of Fig. 3a). This corresponds to a positive magnetoresistance of $(\rho_{8T}-\rho_0)/\rho_0$ = 27.7 % at 2 K. For $BaFe_{1.84}Co_{0.16}As_2$, an abrupt drop in $\rho_{ab}$ is observed below 22 K. The onset transition temperature for 90 % fixed percentage of the normal-state value is $T_c^{onset}$ = 22 K. The transition width in zero field is $\Delta T_c = T_c$ (90%) - $T_c$ (10%) = 0.6 K. The $\Delta T_c$ value is much smaller than that reported for $LaFe_{0.92}Co_{0.08}AsO$ (2.3 K) [15], and for $LaFeAsO_{0.89}F_{0.11}$ (4.5 K) [2]. The resistive transition shifts to lower temperatures by applying 8 Tesla, and transition width becomes wider (1.3 K), characteristic of type-II superconductivity (Fig. 3a, bottom inset). $BaCo_2As_2$ and $BaNi_2As_2$ show metallic behavior (Fig. 3b). At room temperature these materials are more conductive ($\rho_{ab} \approx 0.07$ m$\Omega$ cm) than $BaFe_2As_2$. The residual-resistivity ratio RRR is ~ 8 for each, illustrating good crystal quality. For $BaCo_2As_2$, $\rho_{ab}$ decreases with decreasing temperature, varying approximately linearly with $T^2$ below ~ 60 K (bottom inset of Fig. 3b). By fitting this data using $\rho_{ab}(T) = \rho_{ab}(0\ K) + AT^2$, we obtain the residual resistivity of 5.7 μΩ cm, and $A = 2.2 \times 10^{-3}$ μΩ cm/K$^2$. Comparison with the Sommerfeld coefficient below, gives a Kadowaki-Woods relation $A/\gamma^2 = 0.5 \times 10^{-5}$ μΩ cm mol-Co$^2$ K$^2$/mJ$^2$ consistent with attributing the temperature dependence of the resistivity to electron-electron scattering. For $BaNi_2As_2$, there is a sharp increase at $T_0$ in ρ followed by a continuous decrease with decreasing temperature. The superconductivity downturn in $\rho_{ab}$ is at ~ 1.5 K (top inset of Fig. 3b).

Specific heat data, $C_p(T)$, were also obtained using the PPMS, via the relaxation method below 200 K. For $BaFe_2As_2$ (Fig. 4a), specific heat gives a broad anomaly, peaking at 132(1) K. For $BaFe_{1.84}Co_{0.16}As_2$ (inset of Fig. 4a), there is a specific heat jump below $T_c$ = 22 K, confirming bulk superconductivity. There are no specific heat features for $BaCo_2As_2$ (Fig. 4b) suggesting no phase transition up to 200 K, while for $BaNi_2As_2$ there is a sharp, pronounced transition at $T_0$ = 132 K (Fig. 4b). For $BaNi_2As_2$, the temperature dependence of C/T in several fields for H//*ab* is shown in Fig. 4b, inset. In zero-field, there is a sharp anomaly at $T_c$ = 0.69 K with a jump ΔC = 12.6 mJ/(K mol). Taking the value of the Sommerfeld coefficient at $T_c$, this gives a ratio of $\Delta C/\gamma_n T_c = 1.38$, roughly comparable to a value of 1.43 predicted by weak-coupling BCS theory and confirms bulk superconductivity. As the magnetic field increases, the transition becomes broader and shifts to lower temperatures. The upper critical field is $H_{c2}(0)$ = 0.069 Tesla and the coherence length is $\xi_{GL}(0) \approx 691$ Å, as described in ref. [10]. Below ~ 7 K, C/T has a linear $T^2$ dependence for $BaT_2As_2$ (Fig. 4c). The fitted Sommerfeld coefficient, γ, for $BaFe_2As_2$ is 6.1 mJ/(K$^2$mol) [or 3.0 mJ/(K$^2$mol Fe)]. $BaCo_2As_2$ gives γ = 41.6 mJ/(K$^2$mol) [or 20.8 mJ/(K$^2$mol Co)], consistent with a Fermi liquid plus phonon contribution. The Wilson ratio $R_w = \pi^2 k_B^2 \chi/(3\mu_B^2\gamma)$ for $BaCo_2As_2$ is ~ 7 from $\chi_{ab}$ and ~ 10 from $\chi_c$ (measured at 2 K). These values well exceed unity for a free electron system and indicate that the system is close to ferromagnetism [9]. For $BaNi_2As_2$ the normal state electronic specific heat coefficient, $\gamma_n$, is 13.2 mJ/(K$^2$mol).

First principles calculations for the layered Fe-As superconductors show that the electronic states near the Fermi energy, i.e. those electronic states that are involved in superconductivity, are predominantly derived from Fe *d* states, and that there is only modest hybridization between the Fe *d* and As *p* states [16]. The electronic structures of the corresponding Co and Ni compounds are very similar in a rigid band sense. Specifically, if one compares the electronic structures of $BaFe_2As_2$ [17] $BaCo_2As_2$ [8, 9] and $BaNi_2As_2$ [18, 19] one finds that the band structures and density of states are closely related between the different compounds, the main difference being that the Fermi level is shifted according to the electron count. This is important because it means that the alloys will remain metallic with only moderate alloy scattering. Supercell calculations, motivated by our discovery of superconductivity in $Ba(Fe,Co)_2As_2$ [8] showed that this is in fact the case. Fig. 5 shows projections of the electronic density of states onto the Fe sites and the Co site in a supercell of composition $BaFe_{1.75}Co_{0.25}As_2$, following Ref. [8]. As may be seen, the shapes of the projections onto Fe and Co are remarkably similar, consistent with modest scattering and rigid band alloy behavior. Calculations for $BaMn_2As_2$ show different behavior, however [20]. In this compound, there is strong spin dependent hybridization between Mn and As. This implies that Mn substitution in $BaFe_2As_2$ or the other compounds would likely lead to strong scattering and carrier localization.

## 3. Conclusions

The shrinking of the *c*-lattice parameter is observed from T = Fe to Ni in $BaT_2As_2$, at room temperature. For $BaFe_2As_2$, the tetragonal to orthorhombic *Fmmm* distortion also involves a second-order-like spin-density-wave magnetic transition below $T_N$. In contrast to this broader transition, for $BaNi_2As_2$ there is a sharp, first-order, pronounced feature at $T_0$ = 131 K, associated with a reduction of the lattice symmetry from tetragonal to triclinic $P\bar{1}$. The superconducting critical temperature for $BaNi_2As_2$ is at $T_c$ = 0.69 K. $BaCo_2As_2$, midway between $BaFe_2As_2$ ($d^6$) and $BaNi_2As_2$ ($d^8$), shows no evidence of structural or magnetic transitions. $BaCo_2As_2$ shows evidence of strong fluctuations and is predicted to be in close proximity to a quantum critical point.

The bulk nature of the superconductivity for 8.0(5) % Co-doped $BaFe_2As_2$ single crystal is confirmed by narrow transition widths in ρ(T), χ(T), and also the anomaly in C(T). Based on band structure calculations, the superconductivity of $BaFe_2As_2$ appears to be different from $BaNi_2As_2$. The Ni-phase can be understood within the context of conventional electron-phonon theory, while the high $T_c$ Fe superconductor is not described in this way. Supercell calculations show that $BaFe_2As_2$ will remain metallic with only moderate Co-alloy scattering. Thus the $BaT_2As_2$ compounds, T = Fe, Co, and Ni, while all metallic, show a remarkably wide range of interesting properties, including both conventional and unconventional superconductivity, spin-density-wave antiferromagnetism, and proximity to ferromagnetism. Furthermore, results so far indicate that these materials can be continuously tuned between these states using alloying of these three T elements. This provides an avenue for exploring in detail the interplay between the different states.

## Acknowledgement

Research sponsored by the Division of Materials Science and Engineering, Office of Basic Energy Sciences. Part of this research was performed by Eugene P. Wigner Fellows at ORNL. Work at Los Alamos was performed under the auspices of the U. S. Department of Energy.

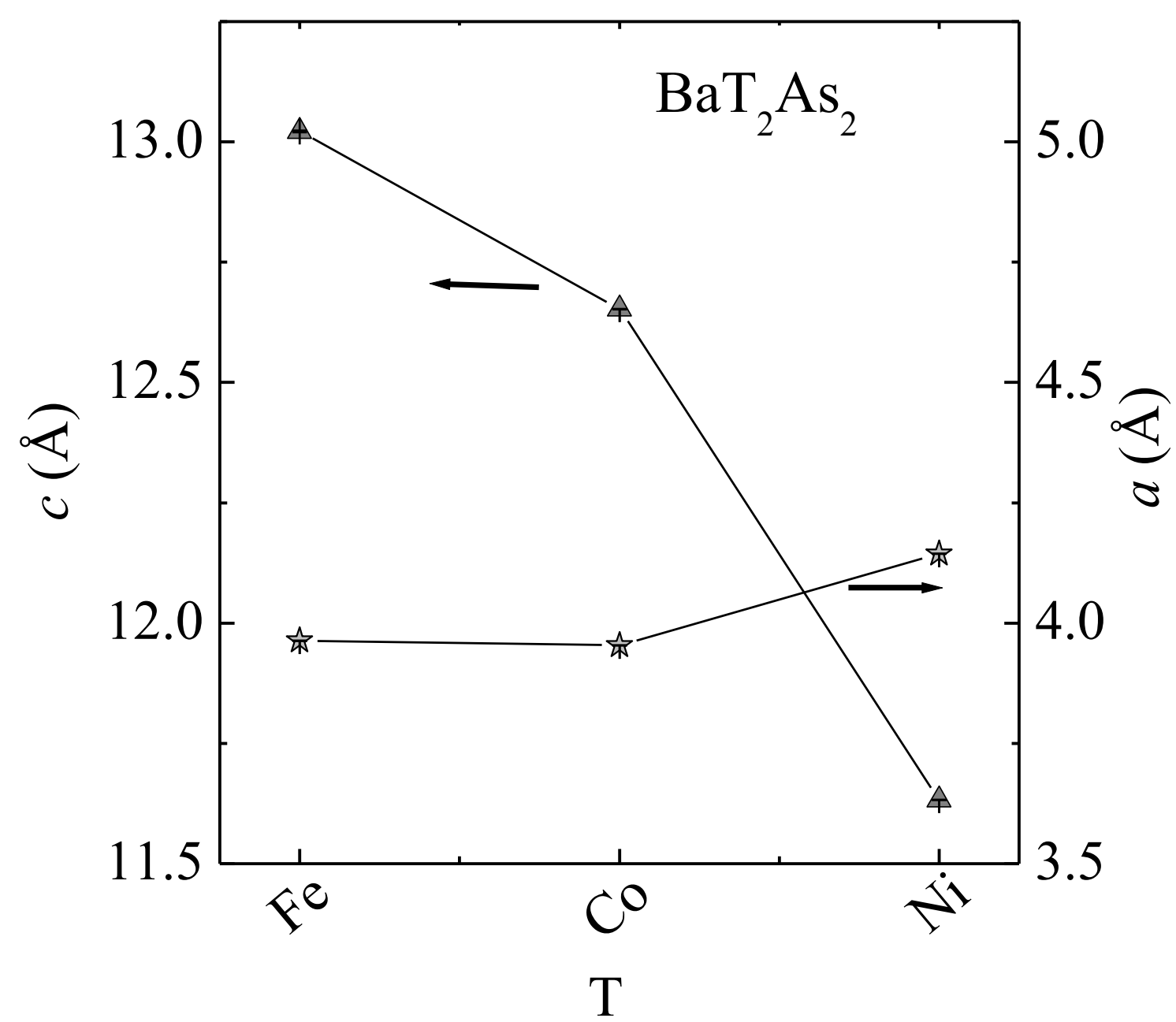

Fig. 1: Room temperature lattice constants for $BaT_2As_2$ with T = Fe, Co, and Ni, refined from x-ray powder diffraction data.

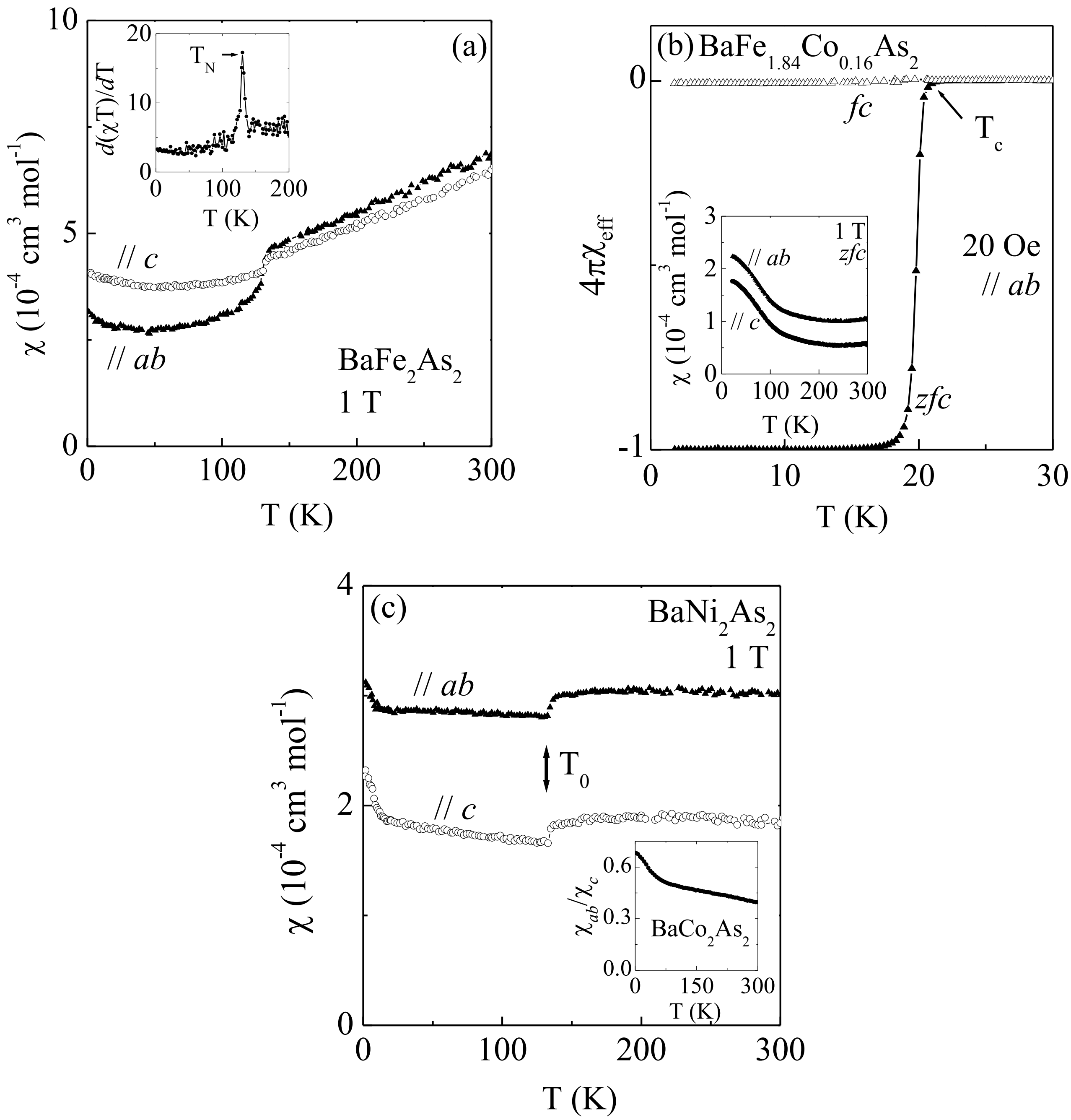

Fig. 2: (a) For $BaFe_2As_2$ and along the two crystallographic directions, the temperature dependence of magnetic susceptibility in zero-field cooled (*zfc*) forms measured at 1 T. The inset shows the average susceptibility, $(2\chi_{ab}+\chi_c)/3$, in the form of $d(\chi T)/dT$ peaking at $T_N$. (b) For $BaFe_{1.84}Co_{0.16}As_2$ in 20 Oe along *ab*-plane, the temperature dependence of susceptibility in *zfc* and field-cooled (*fc*) forms below 30 K, assuming χ value of the perfect diamagnetism. The inset is the χ(T) in *zfc* forms along in 1 T. (c) For $BaNi_2As_2$ and along the two crystallographic directions, the *zfc* temperature dependence of magnetic susceptibility in 1 T. Inset of (c) is the temperature dependence of $\chi_{ab}/\chi_c$ for $BaCo_2As_2$.

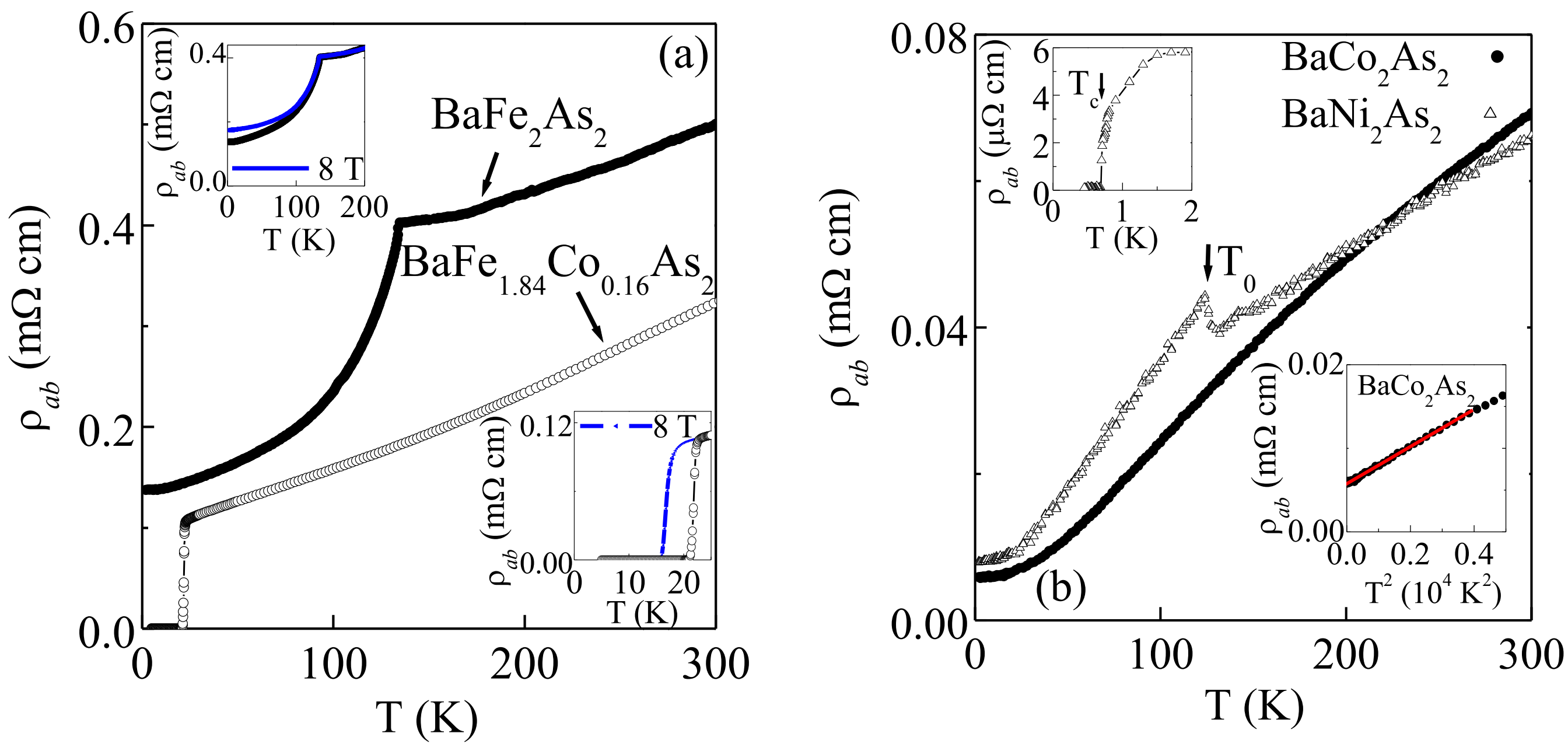

Fig. 3. (Color online) Temperature dependence of resistivity for $BaT_2As_2$ measured in *ab*-plane. (a) $\rho_{ab}$(T) for $BaFe_2As_2$ and $BaFe_{1.84}Co_{0.16}As_2$. Top inset is the behavior measured for $BaFe_2As_2$ at 0 T and 8 T (blue line). Bottom inset is the behavior of $BaFe_{1.84}Co_{0.16}As_2$ at 0 T and 8 T (blue line). (b) $\rho_{ab}$(T) for $BaCo_2As_2$ and $BaNi_2As_2$. Top inset is ρ(T) below 2 K for $BaNi_2As_2$. Bottom inset depicts ρ($T^2$) for $BaCo_2As_2$ between 1.8 K and 70 K, and linear fit below ~ 60 K.

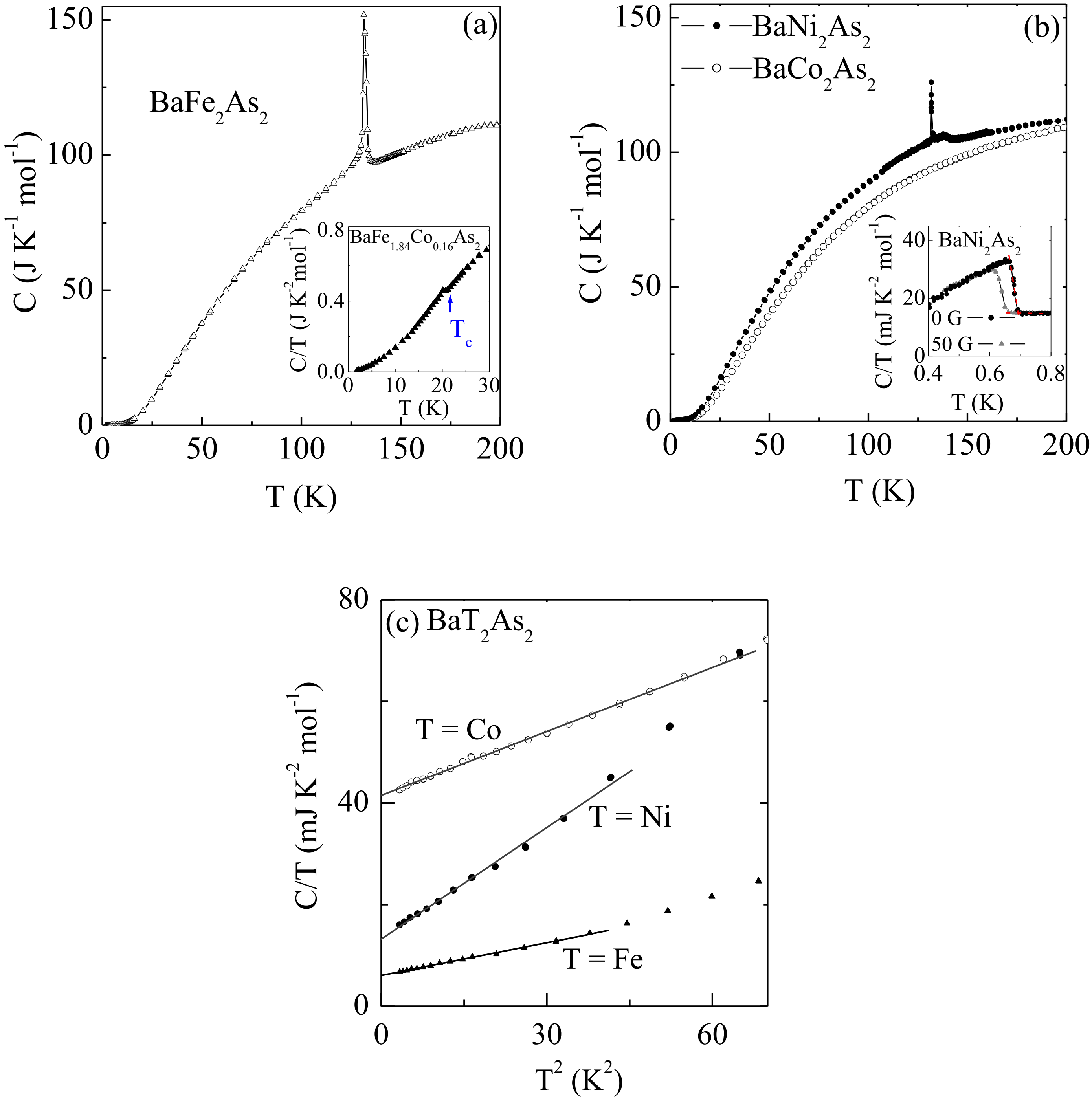

Fig. 4: (Color online) Temperature dependence of the specific heat between 1.8 K and 200 K for (a) $BaFe_2As_2$, and (b) $BaCo_2As_2$ and $BaNi_2As_2$. The inset of (a) is the C/T vs. $T^2$ for $BaFe_{1.84}Co_{0.16}As_2$, depicting anomaly below ~ $T_c$. The inset of (b) is C/T vs. T for $BaNi_2As_2$ below 0.8 K at 0 and 50 G. The $T_c$ was found using the onset criterion demonstrated by the dashed line for the zero-field data. (c) C/T vs. $T^2$ for $BaT_2As_2$ (T = Fe, Co, Ni) and a linear fit between 1.8 K and ~ 7 K.

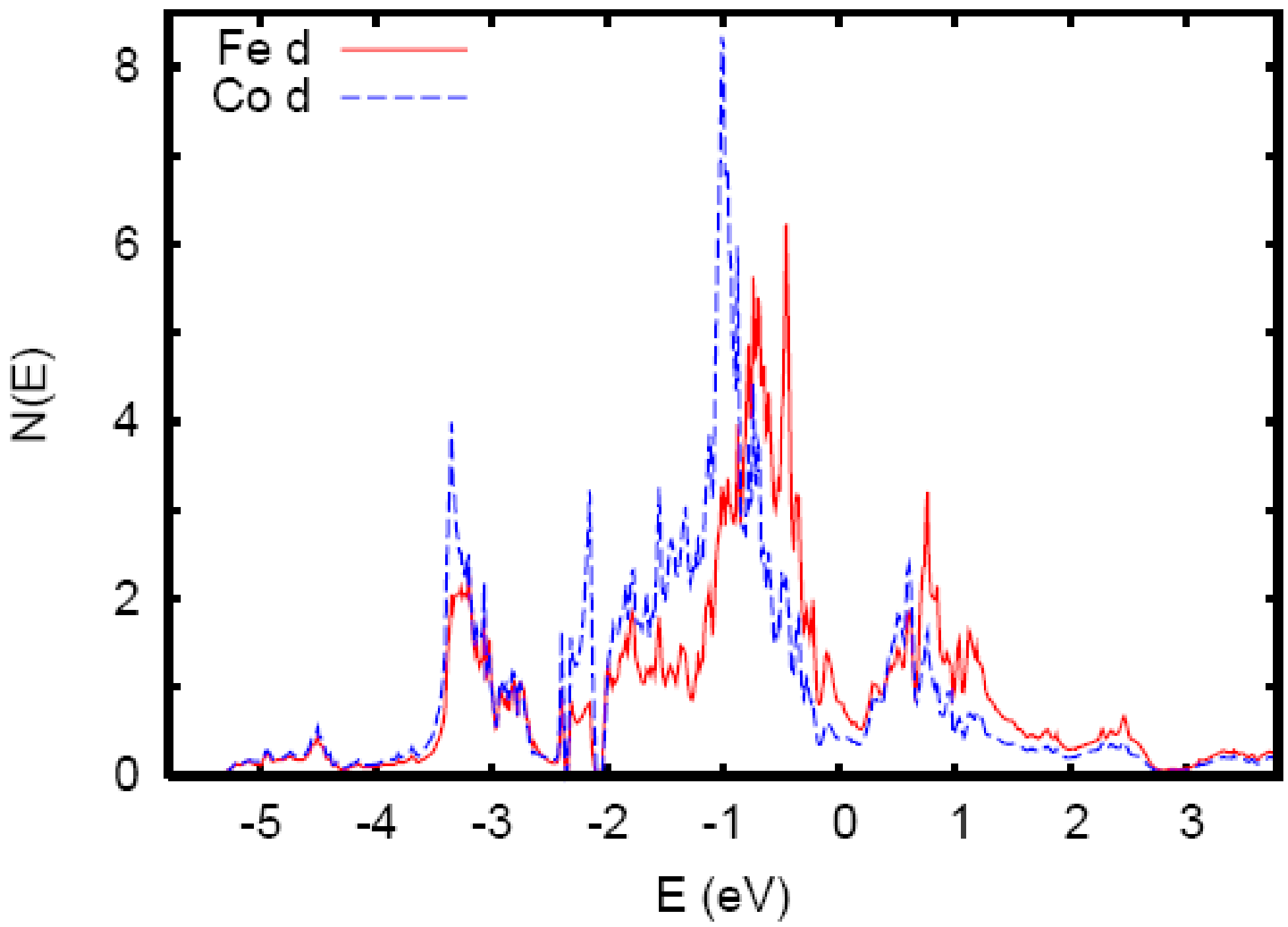

Fig. 5: (Color online): Projected electronic density of states on a per atom basis for a supercell of composition $BaFe_{1.75}Co_{0.25}As_2$ showing that the Fe and Co have similar density of states indicating moderate scattering and near rigid band behavior following Ref. [8].